\documentclass[reprint,
superscriptaddress,
 amsmath,amssymb,
 aps,
pre,
]{revtex4-1}

\usepackage{xcolor}
\usepackage{graphicx}
\usepackage{dcolumn}
\usepackage{bm}

\begin{document}


\title{Relativistic Nature of Carriers: Origin of Electron-Hole Conduction Asymmetry in Monolayer Graphene}

\author{Pawan Kumar Srivastava}
\affiliation{School of Physical Sciences, Jawaharlal Nehru University, New Delhi-110067, India}
\author{Swasti Arya}
\affiliation{Department of Physics, Shiv Nadar University, Gautam Buddha Nagar, Uttar Pradesh-201314, India }%
\author{Santosh Kumar}
\email{skumar.physics@gmail.com}
\affiliation{Department of Physics, Shiv Nadar University, Gautam Buddha Nagar, Uttar Pradesh-201314, India }%
\author{Subhasis Ghosh}
\email{subhasis.ghosh.jnu@gmail.com}
 \affiliation{School of Physical Sciences, Jawaharlal Nehru University, New Delhi-110067, India}
 \affiliation{Department of Physics, Shiv Nadar University, Gautam Buddha Nagar, Uttar Pradesh-201314, India }%


\begin{abstract}
We report electron-hole conduction asymmetry in monolayer graphene. Previously, it has been claimed that electron-hole conduction asymmetry is due to  imbalanced carrier injection from metallic electrodes. Here, we show that metallic contacts have negligible impact on asymmetric conduction and may be either sample or device-dependent phenomena. Electrical measurements show that monolayer graphene based devices exhibit suppressed electron conduction compared to hole conduction  due to presence of donor impurities  which scatter electrons more efficiently. This can be explained by the relativistic nature of charge carriers in graphene monolayer and can be reconciled with the fact that in a relativistic quantum system transport cross section does depend on the sign of scattering potential in contrast to a non-relativistic quantum system.  
\end{abstract}

\pacs{Valid PACS appear here}
\maketitle



There has been much progress in the understanding of fabrication and technology of graphene based devices~\cite{Novoselov2004,Bolotin2008,Avouris2007}. However, there are several issues to be addressed to exploit the high mobility exhibited by monolayer graphene (MLG)~\cite{Lee2014}. One such important issue is electron-hole ($e$-$h$) conduction asymmetry which has been attributed to imbalanced charge injection from the metal electrodes~\cite{Farmer2009,Lopez2010}. It has been claimed~\cite{Farmer2009} that metallic electrodes in graphene based devices create misalignment of neutrality points between electrode and channel, resulting in $e$-$h$ conduction asymmetry. Whereas other groups~\cite{Hannes2011,Huard2008,Xia2011} have claimed that metallic contact-induced electrostatic potential fluctuations is responsible for the $e$-$h$ conduction asymmetry in graphene devices. In these reports, it has been claimed that $e$-$h$ conduction asymmetry is entirely of extrinsic origin. However, there are several reports which showed metallic contact induced doping in graphene without $e$-$h$ asymmetry~\cite{Xia2011,Lopez2012,Leong2014,Laitinen2016}. All these findings have been either interpreted with opposite conclusions or claimed extrinsic mechanism as origin behind $e$-$h$ asymmetry. In view of these results it appears that there are three issues to be resolved: (i) whether metal contacts are responsible for $e$-$h$ conduction asymmetry, (ii) whether it is due to metal induced doping or, (iii) whether $e$-$h$ conduction asymmetry is extrinsic or intrinsic. The objectives laid down in this communication are two-fold. The first objective is to investigate the effect of metallic contacts on asymmetric conduction in MLG based field effect transistors (FETs). We have chosen different metallic electrodes (source/drain) for MLG based devices. In contrast to previous findings, no significant difference in the conduction asymmetry has been observed in our devices with three different metallic electrodes. Hence, the second objective is to probe the factor responsible for asymmetric conduction. We have used two types of graphene monolayers for this study: undoped almost-prestine graphene (type I), and doped graphene (type II). Transfer characteristics, i.e. variation of current between source and drain with gate voltage, in type I MLG based FET always shows symmetry around the Dirac point. On the other hand, type II devices show substantial asymmetry around the Dirac point. Despite different metallic electrodes, these devices do not show any deviation from symmetric/asymmetric conduction around Dirac point suggesting negligible effect of metal induced asymmetry and/or metal induced doping. Moreover, correlation between our experimental findings, i.e. temperature dependence of electron/hole mobility and Raman spectroscopic results with a theoretical model proposed by Novikov~\cite{Novikov2007a,Novikov2007b}, corroborates that difference in scattering cross section due to relativistic nature of charge carriers in graphene is responsible for the asymmetric conduction. It is to be noted that this asymmetry should not be observed in systems in which electrons or holes are governed by non-relativistic quantum mechanics. In case of systems in which electrons or holes are governed by relativistic quantum mechanics, scattering cross section depends on the sign of the scattering potential (donor or acceptor) or sign of the carriers (electrons or holes). In this scenario, the presence of donor/acceptor impurities which scatter electrons/holes more efficiently results in asymmetric conduction around the Dirac point. Particularly in our case, it is the presence of donor impurities which scatter electrons more efficiently as compared to holes, thereby resulting in suppressed electron conduction. In view of our findings, we argue that relativistic nature of carriers should always be taken into account while addressing $e$-$h$ conduction asymmetry in MLG. 

MLG used in this study were grown by chemical exfoliation method. We have used various organic solvents with varying dielectric constant such as toluene, chlorobenzene, acetone, $N$-Dimethylformamide (DMF), and propylene carbonate (PC) for exfoliation. Motivation behind choosing different solvents is to selectively grow MLG with- and without defects. It has been described in our previous report~\cite{Srivastava2013} that using polar and non-polar solvents for exfoliation, MLG can be grown with and without defects, respectively. Raman spectroscopy was performed using WITec GmbH Raman microscope with excitation wavelength of 532 nm. FETs were fabricated on SiO$_2$ (300 nm)/Si ($n^{++}$) substrates in bottom gate configuration. Most of devices contain monolayer graphene flakes which were 15 $\mu$m--25 $\mu$m long and 2 $\mu$m--6 $\mu$m wide with lithographically defined top metallic (Au/Cu/Al; each 45 nm) electrodes (source/drain) with aspect ratio (W/L) of~3. All electrical measurements were performed in vacuum (10$^{-3}$ mbar).

\begin{figure}[t]
\centering
\includegraphics[width=7.5cm]{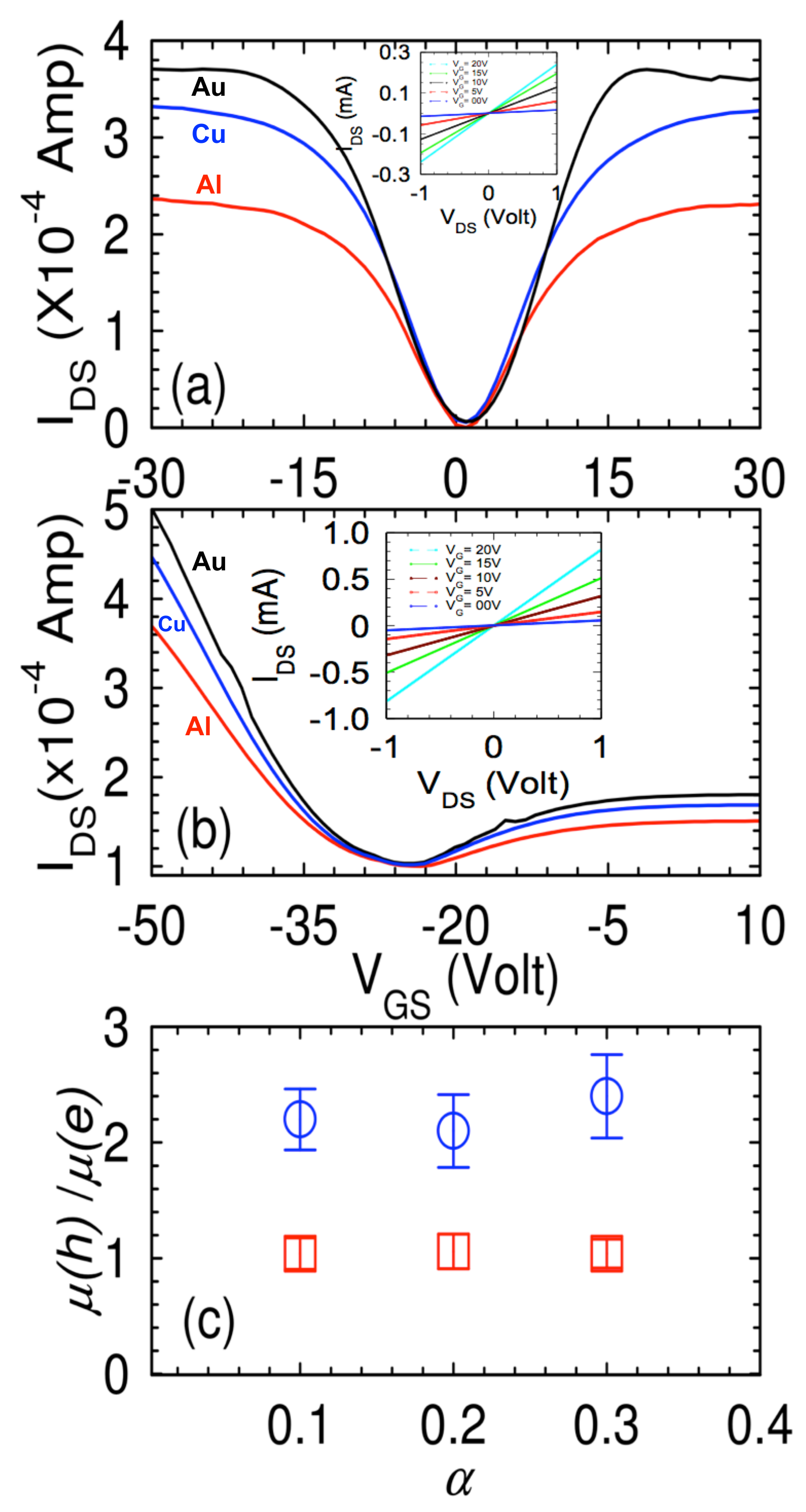}
\caption{Transfer characteristics of FET based on (a) type-I  and (b) type-II MLG, using different metals (Al, Cu, Au) as source-drain electrodes. In all cases, source to drain voltage (VDS) was fixed at 0.1V. (c) Evolution of asymmetric parameter, is plotted against metal (Al, Cu and Au, respectively) work functions. Asymmetric parameter is independent of metallic work functions indicating negligible doping of MLG with metal electrodes.}
\end{figure}

Figure 1 summarizes electrical characteristics of FETs based on type I and type II MLG. In Fig. 1(a), transfer characteristics of type I devices show Dirac point near zero gate bias and symmetric conduction around Dirac point, i.e. similar hole and electron mobilities ($\mu_h$ and $\mu_e$) of $\sim$10,000 cm$^2$/Vs with $\mu_h/\mu_e \sim 1$. Hence, conduction is symmetric when graphene is almost pristine~\cite{SG2015,Giovannetti2008}. In contrast, in Fig. 1(b) transfer characteristics of type II devices show asymmetric conduction around Dirac point with $\mu_e$ and $\mu_h$ of $\sim$ 2600 cm$^2$/Vs, 4000 cm$^2$/Vs ($\mu_h/\mu_e \sim 1.54$) for MLG exfoliated in acetone (type II), 1500 cm$^2$/Vs, 3600 cm$^2$/Vs ($\mu_h/\mu_e\sim 2.4$) for MLG exfoliated in DMF (type II), and 830 cm$^2$/Vs, 2760 cm$^2$/Vs ($\mu_h/\mu_e\sim 3.3$) for MLG exfoliated in PC (type II). We would like to emphasize that asymmetric conduction was reproducible in several type II devices varying in the range of $1.5< \mu_h/\mu_e <3.5$. However, no significant asymmetry has been observed in type I devices. Carrier mobility in all devices has been estimated~\cite{SG2015} using the observed values of gate capacitance in each device instead of using a fixed value of gate capacitance of 11.5nF cm$^{-2}$ corresponding to 300 nm of SiO$_2$ on Si, using the relation $dI_{DS}/dV_{GS} = W \mu C_g V_{DS}/L$, where, $C_g$ is gate capacitance, $W$ is the channel width, $L$ is the channel length, and $I_{DS}$ and $V_{DS}$ denote source-drain current and source-drain voltage, respectively. Moreover, the carrier density for these graphene devices has been obtained using $n = C_g V_D/e$, where $V_D$ is position of Dirac point, and falls in the range of $7.4\times10^{11}$cm$^{-2}$ to $5.5\times10^{12}$cm$^{-2}$.

We should emphasize that in our measurements the contact resistance has negligible impact on the measured mobility. This can be inferred from the observation that the electron and hole mobilities in devices fabricated using different metallic electrodes are almost same. We have also found that $I_{DS}$ varies linearly with $V_{DS}$ suggesting that the contacts between metal and graphene do not limit the current. Moreover, we observed that the total device resistance decreased by a factor of about 12 on changing the gate bias from 0V to 20V (see insets in Fig. 1). Considering that the total resistance is given by $R_T=R_C+R_S$, where $R_C$ and $R_S$ are contact and sheet resistances, respectively, such a huge reduction in resistance can only be attributed to the decrease in $R_S$, thereby ruling out any significant impact of contact resistance on mobility~\cite{Xia2011,Bartolomeo2013}. 

The transfer characteristics of type I and type II devices, using Al, Cu and Au as source-drain electrodes, are also shown in Fig 1. It can be clearly seen in the characteristics of type I devices that despite different metallic electrodes Dirac point is always positioned near 0V and transfer characteristics show symmetry ($\mu_h/\mu_e  \sim 1$) around Dirac point. Type II devices with different metallic contacts also show similar behaviour. It can be clearly seen in the characteristics of type II devices that despite different metallic electrodes Dirac point is positioned at  $\sim -26$V and transfer characteristics show strong asymmetry ($1.5\lesssim\mu_h/\mu_e\lesssim2.5$) around Dirac point. Evaluation of $\mu_h/\mu_e$ for different metallic electrodes is summarized in Fig. 1(c) for type I and type II devices. Clearly, we do not see any dependence of symmetry or asymmetry on different metallic electrodes (Al, Cu and Au). It is clear that, by varying the metallic contacts, individual characteristics of type I and type II devices have not changed. We would like to emphasise that discrepancy between our results and previous reports~\cite{Farmer2009,Lopez2010,Hannes2011,Das2008} regarding the role of top metallic contacts in determining the conduction asymmetry could be due to relatively high field effect electron/hole mobility in our devices. For instance, Ref.~\cite{Farmer2009}, which suggests imbalanced carrier injection from metallic electrodes is responsible for asymmetric conduction, presents data on devices with a relatively low mobility of 100 cm$^2$/Vs which is at least one order of magnitude smaller than the mobility in our devices. We suspect that the effect of metallic contacts could be vital if we deal with low mobility MLG samples with high concentration of defects. Now, it opens up a debate about the factor responsible for conduction asymmetry in type II devices.

\begin{figure}[t]
\centering
\includegraphics[width=7cm]{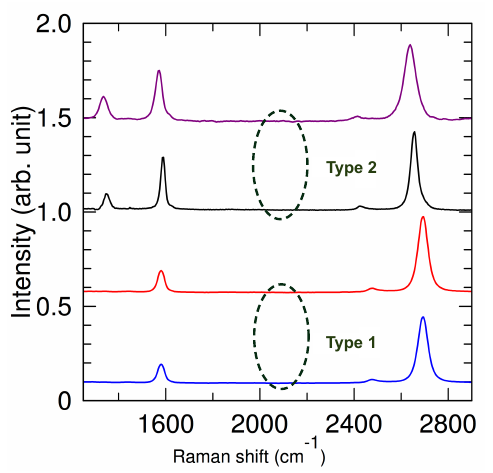}
\caption{Raman spectra of (from bottom to top) type-I and type-II MLG layers.  In type-I MLG layers,  Raman D peak is absent  whereas, in type-II MLG layers,   D peak is present.  Monotonic red shift in 2D Raman peak positions which is an indication of doping can be seen in case of type-II MLG layers. Individual spectrum is shifted on Y-axis for better clarity.}
\end{figure}
\begin{figure}[ht]
\centering
\includegraphics[width=8cm]{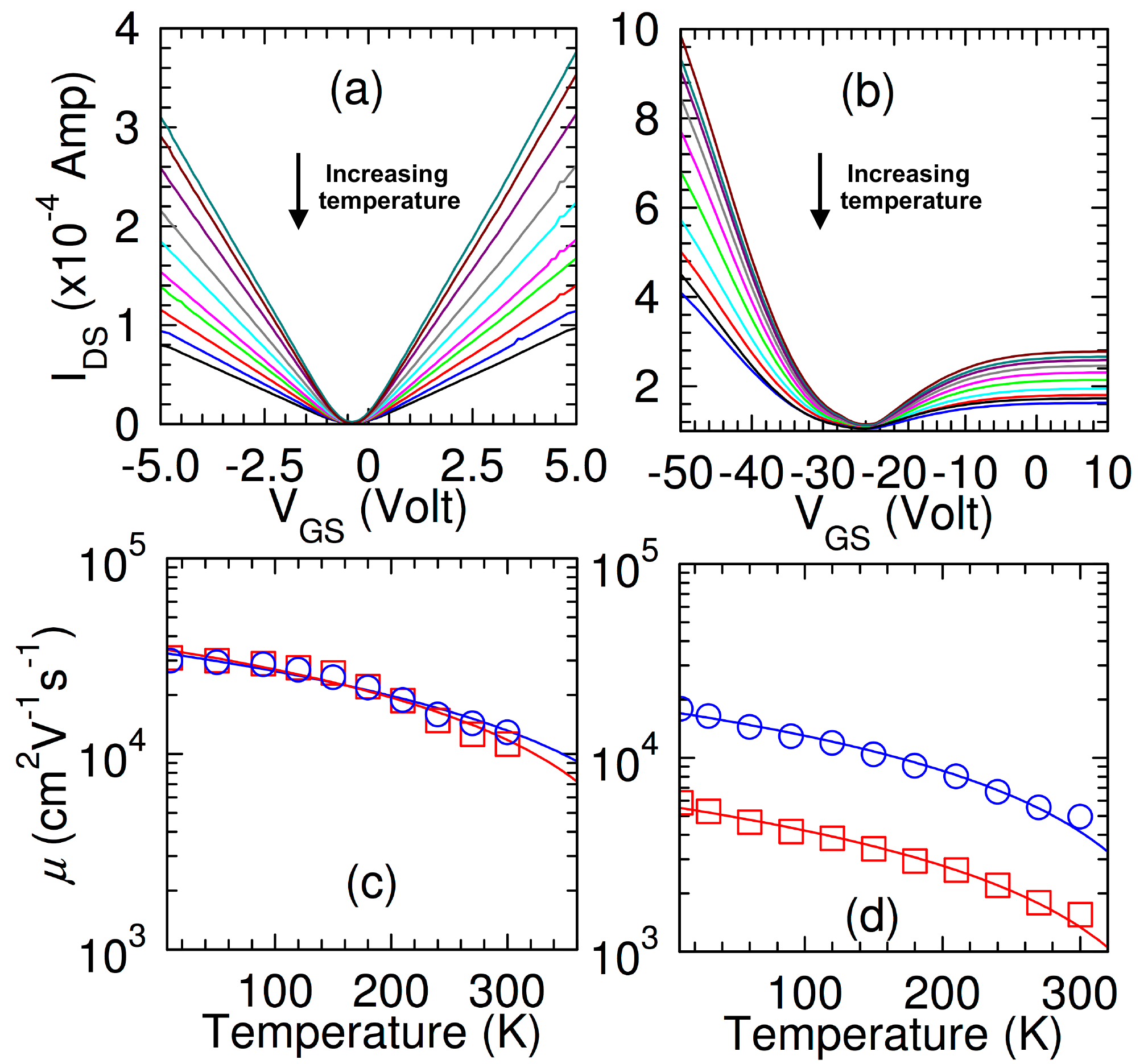}
\caption{Temperature dependence of transfer characteristics of FET based on  (a) type-I  and (b) type-II MLG layers. Temperature dependence of carrier mobility in  (c) type-I and (d) type-II MLG layers, extracted from data given in (a) and (b). Hole and electron mobility ratio as a measure of conduction asymmetry at 300K/10K has been found to be 1.02/1.04 and 3.0/3.2 in case of  type-I and type-II MLG layers,  respectively}
\end{figure}

Figure 2 summarizes the Raman spectra of type I and type II MLG. For type I MLG, Raman G and 2D peak were observed at 1580 cm$^{-1}$ and 2695 cm$^{-1}$ with almost no D band, indicating the absence of any in-plane or out of plane disorder in graphene sample~\cite{SG2015}. In contrast, two significant differences can be observed in Raman spectra of type II MLGs, (i) presence of strong D band and (ii) red shift in G and 2D peak positions which are at 1570 cm$^{-1}$ and 2660 cm$^{-1}$, respectively. As illustrated in Fig. 1, type II devices exhibit Dirac point at higher negative back gate bias ($\sim -26$V) indicating $n$-type doping which is corroborated by the Raman signatures~\cite{Srivastava2013}. In our previous work~\cite{Srivastava2013,SG2015}, we have shown that Raman D peaks in doped MLG (type II) are due to functionalization with organic molecules on graphitic surface as evidenced~\cite{Huttmann2015,Wehling2008} by scanning tunneling microscope and infrared spectroscopic measurements~\cite{SG2015}. Moreover, DFT calculations~\cite{Srivastava2015a,Srivastava2015a} also suggest that in case of graphene prepared in polar solvents, molecules are close enough to graphitic surface to have significant impact on Raman D peak. 

In a recent study~\cite{Novikov2007a,Ostrovsky2006}, it has been shown that relativistic nature of carriers in MLG plays pivotal role in determining the exact transport cross section which is quite sensitive to whether the charge carrier is attracted to an impurity or is repelled from it. For example, for a conduction electron the transport cross section is greater when scattering off a donor than off an acceptor. This attraction/repulsion asymmetry is an unanticipated feature for scattering due to the Coulomb potential, and arises due to the underlying relativistic nature of the problem. Indeed, such an asymmetry is absent in the nonrelativistic scattering off a Coulomb potential in two~\cite{Stern1967} and three~\cite{Landau1977} dimensions. Moreover, this kind of conduction asymmetry is also absent for scattering both off neutral imperfections and off quenched corrugations~\cite{Ostrovsky2006}. Hence, the observation of asymmetry only in doped MLG supports the theoretical prediction outlined in Refs.~\cite{Novikov2007a,Novikov2007b}. It can be emphasized that the presence of donor impurities in doped MLG (type II devices) is responsible for such suppressed electron conduction. To settle down this issue, we have first carried out temperature dependent electrical measurements. Figure 3 shows the temperature dependent transfer characteristics of one of the type I and type II devices. Dirac point in type I device remains close to zero gate bias with $\mu_h/\mu_e \sim 1.04$ at room temperature and $\mu_h/\mu_e\sim 1.02$ at low temperature, indicating symmetric conduction and pristine behavior of graphene exfoliated in toluene. Similar characteristics has also been observed for graphene exfoliated in chlorobenzene (type I). Similarly, Dirac point and $e$-$h$ conduction asymmetry in graphene exfoliated in DMF (type II) do not change in the measured temperature range and estimated to be $\mu_h/\mu_e \sim 3.2$ and 3.0 at room temperature and low temperature, respectively. The MLG exfoliated in acetone (type II) exhibits similar behaviour and we have estimated $\mu_h/\mu_e \sim 1.5$ at room temperature as well as at low temperature. In case of type II devices, existence of similar asymmetry i.e. $\mu_h/\mu_e$ values in measured temperature range suggests that carrier scattering responsible for asymmetric conduction is indeed due to scattering off charged impurities~\cite{SG2015,Chen2008,Zhu2009}. To reconcile our results, we have compared our experimental results with the theoretical model proposed in Refs.~\cite{Novikov2007a,Novikov2007b}.

Consider the scattering of a particle off the Coulomb potential 
\begin{equation}
U(r)=-Ze_*^2/r=-\hbar v \alpha_0/r.
\end{equation}
Here, $Z$ is the impurity valance , and $e_*^2=2e^2/(\varepsilon+1)$, with $\varepsilon$ being the dielectric constant of a substrate. $\alpha_0 (=Ze_*^2/(\hbar v))$ is the dimensionless impurity strength with $v$ representing the Fermi velocity. The effect of interactions between carriers in graphene diminishes the value of the impurity strength compared to its value in vacuum. This effective value of the dimensionless impurity strength is denoted by $\alpha$, with $0<\alpha<1/2$. The upper cut-off value of 1/2 corresponds to the subcritical impurity~\cite{Novikov2007a,Novikov2007b}. If $j(=m+1/2)$ be the angular momenta, the scattering phase shifts associated with the the scattering states is given by~\cite{Novikov2007a,Novikov2007b}
\begin{equation}
\delta_j=\frac{1}{2i}\left[\log\frac{je^{i\pi(j-\gamma)}\Gamma(1+\gamma-i\alpha_\varepsilon)}{(\gamma-i\alpha_\varepsilon)\Gamma(1+\gamma+i\alpha_\varepsilon)}\right],
\end{equation}                                                                
where, $\gamma=\sqrt{(j^2-\alpha^2)}$,  $\alpha_\varepsilon=\alpha \text{ sign}(\epsilon)$, and  $\epsilon$ represents the quasiparticle energy.
The main objective is to find out the imbalance in conduction of different type of charge carriers. It turns out that the transport cross section (in the units of carrier wavelength) for a charge carrier is
\begin{equation}
\label{Calp}
C(\alpha)=\frac{2}{\pi}  \sum_{j=1/2}^{\infty}\sin^2 (\delta_{j+1}-\delta_j),
\end{equation}
which is strongly asymmetric with respect to the type of potential seen by the carrier, i.e. attractive or repulsive~\cite{Novikov2007a}. As a consequence, donors scatter electrons with relatively higher cross-section as compared to the holes. Similarly, acceptors scatter holes with a higher cross-section than electrons. The cross-section, as given by Eq.~\eqref{Calp}, appears explicitly in the kinetic equations for the charge carriers which have to be solved to determine the dc conductivity. As shown in~\cite{Novikov2007a}, this dc conductivity comes out to be
\begin{equation}
\label{sigma}
\sigma=\frac{(e^2/h)p}{n_i^+C(-\alpha)+n_i^-C(\alpha)}+\frac{(e^2/h)n}{n_i^-C(\alpha)+n_i^- C(-\alpha)}.
\end{equation}                           	  		
Here $p$ and $n$ are hole and electron densities,  $n_i^+$ and $n_i^-$ are positive (donor) and negative (acceptor) impurity densities, respectively. Therefore, we obtain the hole and electron mobilities to be
\begin{equation}                                  
\mu_h=\frac{(e/h)}{n_i^+C(-\alpha)+n_i^-C(\alpha)},~~~\mu_e=\frac{(e/h)}{n_i^+C(\alpha)+n_i^-C(-\alpha)}.                       
\end{equation}
Consequently, the $e$-$h$ conduction asymmetry can be expressed as the ratio of hole and electron mobilities:
\begin{equation}
\label{ratio}
\frac{\mu_h}{\mu_e}=\frac{n_i^+C(+\alpha)+n_i^{-}C(-\alpha)}{n_i^+C(-\alpha)+n_i^-C(+\alpha)}.
\end{equation}
It is clear from Eq.~\eqref{ratio} that $\alpha$ plays a crucial role in determining conduction asymmetry. We should emphasize that the relativistic formulation is valid only near the Dirac point where the energy dispersion in graphene is linear~\cite{Zhang2005}. It corresponds to linear region of the transfer characteristics which has been used for the mobility calculation in the present work.

\begin{figure}[ht]
\centering
\includegraphics[width=7cm]{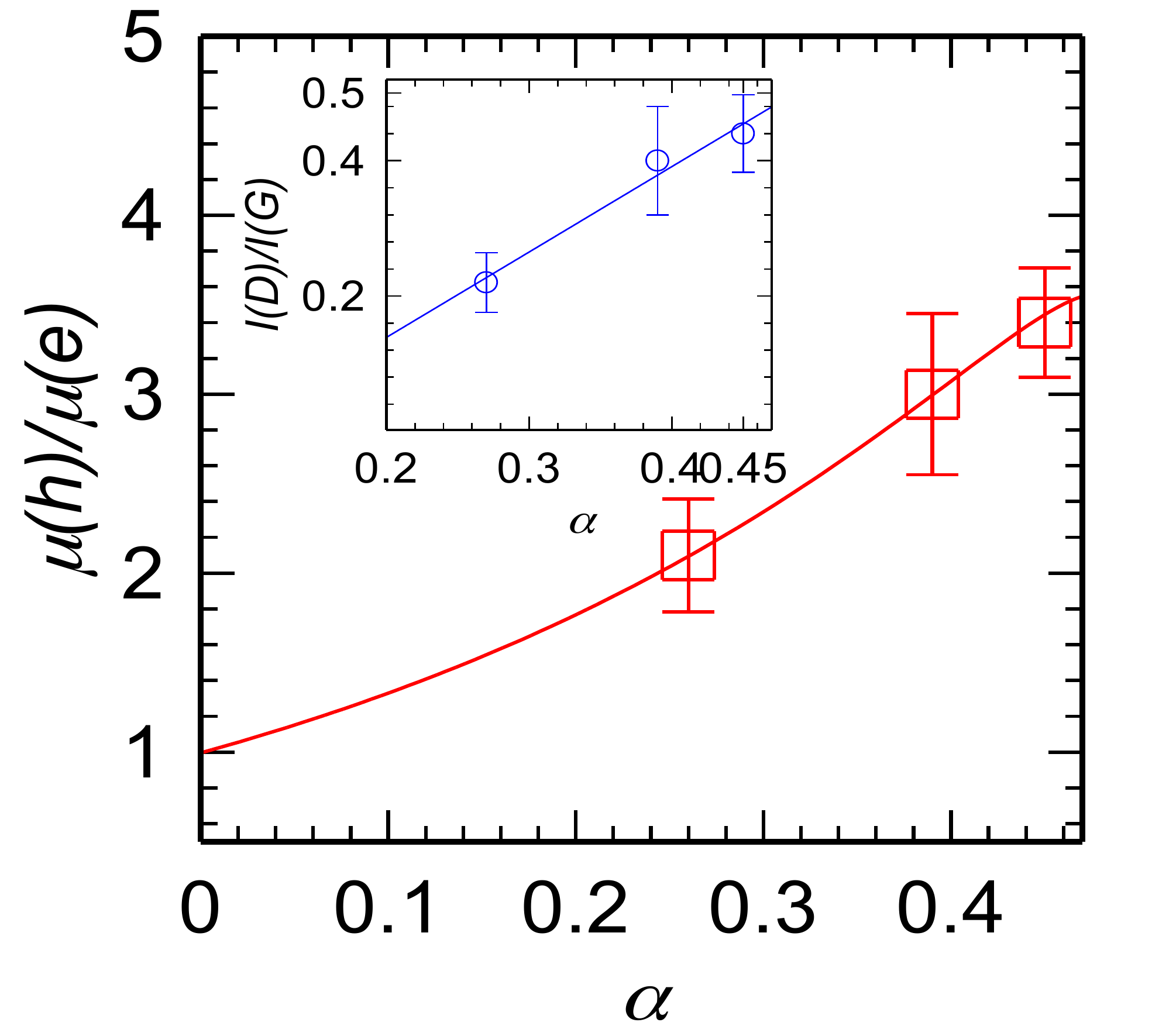}
\caption{$\mu_h/\mu_e$ as a function of impurity strength ($\alpha$) as expressed by Eq.~\eqref{ratio} given in text (solid line). Symbols are the experimentally observed for MLG based devices. Inset shows the  dependence of normalized Raman D peak intensity on the impurity strength ($\alpha$). Y-axis error bars denote uncertainty in $\mu_h/\mu_e$ and $I_D/I_G$ values measured on minimum five samples. }
\end{figure}
For type I devices, which involve exfoliation with toluene, earlier studies~\cite{Kaverzin2011,Srivastava2017} have suggested that the nature of doping is nontrivial. Therefore, we focus only on type II devices and try to correlate the observed asymmetry with the above theoretical model. We use the relation $ \mu n_i \sim 5\times10^{15}(\text{Vs})^{-1}$~\cite{HAS2007,Chen2008a} to obtain the impurity density to be about $n_i \sim 10^{11}-10^{12}\text{cm}^{-2}$. Here $\mu$ is the electron or hole mobility, and $n_i$ is the overall impurity concentration. This range is also consistent with the Dirac point shift of about 30V in the transfer characteristics with gate capacitance of about 10nF/cm$^2$. In the type II devices, since the donor impurity concentration is prevalent we attributed $n_i$ entirely to donors i.e., $n_i^+$. We found that even if we use nonzero $n_i^-$ with concentration one-two order magnitude less than $n_i^+$, then the results do not change significantly. Therefore, we consider $n_i^- = 0$, and then try to fit $n_i^+$ and $\alpha$ value that produce mobility values closest to the experimental values. We found that for type II devices, $\alpha$ falls in the range $0.24\lesssim \alpha \lesssim 0.46$ and lead to mobility values and the corresponding asymmetry-ratios very close to that obtained using the experimental data.  We summarize below the calculations for a typical choice of values (with proper units):\\
\noindent
{\bf Acetone exfoliated graphene:} \\
$~~~~~n_i^+=8\times 10^{11}, n_i^-=0, \alpha=0.26$\\
$\Rightarrow \mu_h\approx3890, \mu_e\approx1970, \mu_h/\mu_e\approx2$
\vspace{0.1cm}\\
{\bf Dimethylformamide exfoliated graphene:} \\
$~~~~~n_i^+=5\times 10^{11}, n_i^-=0, \alpha=0.39$\\
$\Rightarrow \mu_h\approx3090, \mu_e\approx1120, \mu_h/\mu_e\approx2.8$
\vspace{0.1cm}\\
{\bf Propylene Carbonate  exfoliated graphene:} \\
$~~~~~n_i^+=4.4\times 10^{11}, n_i^-=0, \alpha=0.45$\\
$\Rightarrow \mu_h\approx2740, \mu_e\approx865, \mu_h/\mu_e\approx3.2$
\vspace{0.1cm}\\
These mobility values and the ratio fall within the range of values observed in the experiment, as shown in Fig. 4. We have also plotted the normalized Raman D peak intensity $(I_D/I_G)$ as a function of $\alpha$ extracted for type II devices in inset of Fig. 4. We can see that $\alpha$ varies linearly with $I_D/I_G$. The idea behind comparing linear relation of $I_D/I_G$ with $\alpha$ is based on the fact that in our previous publication~\cite{SG2015} we have shown that $I_D/I_G$ evolves linearly with impurity density in graphene devices. In this view, Fig. 4 suggests impurity strength $\alpha$ should also vary linearly with impurity density which is nothing but density of donor impurities in our case. Hence, linear variation of $\alpha$ with $I_D/I_G$ suggests that impurity strength which defines the asymmetric conduction in MLG based devices is directly correlated with strength of donor impurity density in MLG samples. Therefore, it can be concluded that the suppressed electron conduction in type II devices is a consequence of asymmetric scattering cross section for relativistically defined hole and electron transport followed by local donor impurity centers which scatter off electrons more efficiently as compared to holes. 

In conclusion, we have investigated the $e$-$h$ conduction asymmetry in MLG based devices. Our analysis rules out any substantial impact of doping with metallic contacts on conduction asymmetry. In addition to temperature dependent evolution of electron and hole mobility ratio which is a measure of asymmetry, analysis on Raman spectroscopic results support our claim. Our results are in agreement with previous reports~\cite{Novikov2007a,Bai2015} indicating importance of relativistically defined carriers in the description of asymmetric conduction in MLG based devices. 

\acknowledgements
P.K.S. acknowledges CSIR, India, for financial assistance. S. A. thanks Shiv Nadar foundation, Shiv Nadar University, India, for financial assistance. Authors are also grateful to Hitesh Mamgain, WITec GmbH for his help in Raman spectroscopic measurements.

\end{document}